\documentclass[aps,floatfix,reprint,superscriptaddress,nofootinbib]{revtex4-2}

\usepackage{amsmath,amssymb,graphicx,mathtools}
\usepackage{float}

\DeclareMathOperator{\arcsinh}{arcsinh}
\usepackage[T1]{fontenc}

\usepackage{color}
\usepackage{colortbl}
\usepackage{nicefrac}
\usepackage{tabularx}

\usepackage{calligra}

\usepackage{tikz-cd}

\usepackage{multirow}

\usepackage{diagbox}

\usepackage{scalerel}
\usepackage{stackengine,wasysym}

\makeatletter
\DeclareRobustCommand{\cev}[1]{%
  {\mathpalette\do@cev{#1}}%
}
\newcommand{\do@cev}[2]{%
  \vbox{\offinterlineskip
    \sbox\z@{$\m@th#1 x$}%
    \ialign{##\cr
      \hidewidth\reflectbox{$\m@th#1\vec{}\mkern4mu$}\hidewidth\cr
      \noalign{\kern-\ht\z@}
      $\m@th#1#2$\cr
    }%
  }%
}
\makeatother

\usepackage{physics}
\usepackage[mathscr]{euscript}

\usepackage{blkarray}

\newcounter{theoremcounter}
\stepcounter{theoremcounter}

\newcommand{\dee}{\text d}

\newcommand{\rtt}{\mathbb{R}^{2,2}}
\newcommand{\rttz}{\rtt/\mathbb{Z}_2}

\newcommand{\ads}{\textrm{AdS}_3}

\newcommand{\spacetime}{\text{NO-AdS}_3}

\newcommand{\slral}{\mathfrak{sl}(2,\mathbb{R})}
\newcommand{\slr}{\textrm{SL(2, $\mathbb{R}$)}}

\newcommand{\mplr}{\textrm{Mp(2, $\mathbb{R}$)}}

\newcommand{\sorthog}{\textrm{SO}(2,2)}

\begin{document}

\title{Non-Orientable AdS$_3$ and its Holography}
\author{Shivesh Pathak and Lucas Kocia Kovalsky}
\affiliation{Quantum Algorithms and Applications Collaboratory, Sandia National Laboratories, Livermore, California 94550, U.S.A.}
\date{\today }
\begin{abstract}
  Known solutions to three-dimensional gravity with negative cosmological constant so far 
  consist of either $\ads$ or its orbifolds (or orientifolds). We geometrically derive a novel non-orientable AdS$_3$ spacetime that is an orientifold of a spinor double cover of $\ads$, unlike existing solutions. 
  This spacetime's universal cover contains a lightlike instead of spacelike compact direction. We show that its gravity theory possesses a simpler excited state spectrum and exhibits a Chern-Simons/Wess-Zumino-Witt (CS/WZW) correspondence between only one independent CS term and boundary chiral WZW term. 
This leads us to suggest that CS theory is able to more fully describe non-orientable pure gravity theory compared to orientable gravity.
\end{abstract}
\maketitle

\section{Introduction}
\label{sec:intro}

Anti-de Sitter (AdS) spacetime is a useful testbed to probe the features of quantum gravity. Its constant negative curvature yields a uniquely timelike boundary that is able to support a conformal field theory (CFT). According to the AdS/CFT correspondence~\cite{Maldacena99,Aharony00,Hayden07}, bulk AdS string theory is conjectured to correspond to this CFT, which allows its gravity theory to be modelled as an emergent phenomenon from the collective action of quantum fields acting on its boundary: the holographic effect of a lower-dimensional manifold~\cite{Ryu06,Ryu06_2,Headrick07,Brown16,Brown16_2,Susskind16}. 
However, despite much study and progress, the dictionary between CFT properties and bulk AdS spacetime has often been found to be approximate~\cite{Witten98,Gubser98,Banks98,Susskind98,Polchinski99,Giddings99}, and the AdS/CFT correspondence has yet to be fully characterized.

Due to its vanishing Weyl tensor, solutions to three-dimensional gravity with negative cosmological constant must locally possess $\ads$'s isometries~\cite{Banados93}. This is often taken to mean that only $\ads$, $\ads$ orbifolds, or $\ads$ orientifolds are allowed, i.e.~quotients of $\ads$ by discrete $\mathbb Z_n$ or $\mathbb Z$ subgroups of $\text{SO}(2,2)$ or $\text{O}(2,2)$, respectively~\cite{Loran10} (or covers thereof).

The BTZ blackhole spacetime~\cite{Banados92,Banados93} is a well-known example of an $\ads$ quotient spacetime, and is a discrete $\mathbb Z$ quotient of AdS$_3$. 
An example of a cover is the universal cover of $\ads$, which has a time dimension corresponding to $\ads$'s unrolled periodic time.~\footnote{In this paper we refer to ``a single cover of $\ads$'' as $\ads$. When needed, the ``universal cover of $\ads$'' will be referred to explicitly as such.} 
A more relevant example for us is the double (or spinorial) cover of $\ads$'s isometry group $\text{SO}^+(2,2)$: $\text{Spin}^+(2,2) \sim \slr \times \slr$ (a connected group not to be confused with the disconnected $\sorthog$).

Unlike in four dimensions or higher, 
Einstein's equations do not allow for propagating degrees of freedom in three dimensions. This greatly simplifies its gravity theory and may permit its equivalent description solely in terms of topological invariants~\cite{Donnay16}. As a result, a simpler bulk/boundary correspondence can be derived between a pure gauge theory of three-dimensional AdS ($\ads$) and a type of two-dimensional CFT (CFT$_2$) at its boundary. For a spacetime with isometry group $G\times G$, the topological theory is Chern-Simons (CS) with gauge group $G$~\cite{Achucarro86,Witten88}, and the boundary CFT$_2$ is the Wess-Zumino-Witten (WZW) sigma-model defined on the Lie group $G$~\cite{Wess71,Witten83,Witten84,Gawedzki88}. While this CS/WZW correspondence is more formally exact compared to the conjectured AdS/CFT correspondence, for non-compact $G$ groups such as AdS$_3$'s $\text{SO}^+(2,2) \sim (\slr\times \slr)/\mathbb Z_2$, the bulk-boundary CS/WZW suffers from some complications:

Firstly, three-dimensional gravity permits an additional term to the standard Einstein-Hilbert action for orientable spacetimes, comprised of a three-form in addition to the standard two-form~\cite{Deser82,Witten88}. This allows $\slr$ CS actions to consist of linear combinations of independent left-~and right-handed chiral contributions, which should correspond to separate factors of a tensor-product Hilbert space upon quantization. 
Consequently, it becomes unclear how to classically interpret states consisting of a ground state tensored with excited states that are orbifolds of both $\slr$ factors, such as BTZ states~\cite{Witten07}.

Secondly, 
it seems unlikely that CS theory can describe the highly degenerate states of the BTZ blackhole given the theory's limited topological degrees of freedom. This brings into doubt the ability of three-dimensional gravity with negative cosmological constant to be fully equivalent to CS theory even at the perturbative level.

Lastly, non-perturbatively, the gauge field of CS theory may cease to be invertible and the corresponding non-geometrical configurations may need to be included~\cite{Witten07}. 
It is also expected that a sum over inequivalent topologies and an invariance under disconnected diffeomorphisms, such as modular transformations, is necessary for quantum gravity. CS theory seems to only naturally include diffeomorphisms connected to the identity.

It would be useful to find a quotient $\ads$ spacetime whose gravity theory is more clearly and fully described by CS theory. 
If this is possible, it likely requires changing some of the properties of the ground state $\ads$ spacetime under consideration. 
Orientability is an especially compelling property to target; prior work has show that generalizing $\slr$ CS theory to a non-orientable gauge manifold 
results in only a single independent CS term 
instead of two~\cite{Chen14}, which may lead to a simpler Hilbert space upon quantization. 
Compellingly, this generalized CS theory seems to suggest there exists a non-orientable $\ads$ spacetime that is not an orientifold $\ads$ spacetime like existing known solutions. 
Without knowing the geometry of its corresponding non-orientable spacetime, it is unclear how relevant this non-orientable generalization of CS theory is to gravity theory.

Here we will geometrically derive a non-orientable $\ads$ spacetime that is not an orientifold of regular $\ads$, but rather of its double (spinorial) cover, $\text{Spin}^+(2,2)$. We will find that this novel spacetime's universal cover contains a lightlike instead of spacelike compact
direction at its timelike boundary. We will also find that it exhibits a Chern-Simons/Wess-Zumino-Witt (CS/WZW) correspondence between only one independent CS term and chiral WZW term at its boundary, as suggested in~\cite{Chen14}, and that this gravity theory possesses a simpler excited state spectrum without BTZ black holes. This will lead us to suggest that CS theory is able to fully describe non-orientable pure gravity theory in three dimensions perturbatively with negative cosmological constant.

This paper is organized as follows. In section~\ref{sec:construction} we geometrically derive the non-orientable AdS$_3$ ($\spacetime$) spacetime. 
In section~\ref{sec:correspondence} we discuss $\spacetime$'s more exact CS theory and how its CS/WZW correspondence manifests as a simpler relationship between a single CS and WZW term. We discuss the significance of our results in section~\ref{sec:discussion} and conclude in section~\ref{sec:conc}.

\section{Constructing non-orientable $\ads$}
\label{sec:construction}

The bulk action in CS theory produces trivial dynamics for a given three-dimensional spacetime. On the other hand, asymptotic symmetries on the boundary can break bulk gauge invariance and generate an infinite number of global degrees of freedom~\cite{Donnay16}.

Motivated by this, in this section we will derive a non-orientable $\ads$ spacetime by introducing a topologically non-trivial orientifold on a double cover of $\rtt$, which will be inherited by its embedded $\ads$s. 
First, we will review well-known relationships between $\rtt$ and $\ads$ in section~\ref{sec:preamble} to establish our notation. We will then proceed to derive a double cover space and $\mathbb Z_2$ quotient of $\rtt$ in section~\ref{sec:doublecover}. In section~\ref{sec:poincarepatches} we will see how these transformations are inherited by its embedded $\ads$s.

\subsection{Preliminaries}
\label{sec:preamble}

\begin{figure}
  \centering
  \includegraphics[width=0.5\textwidth]{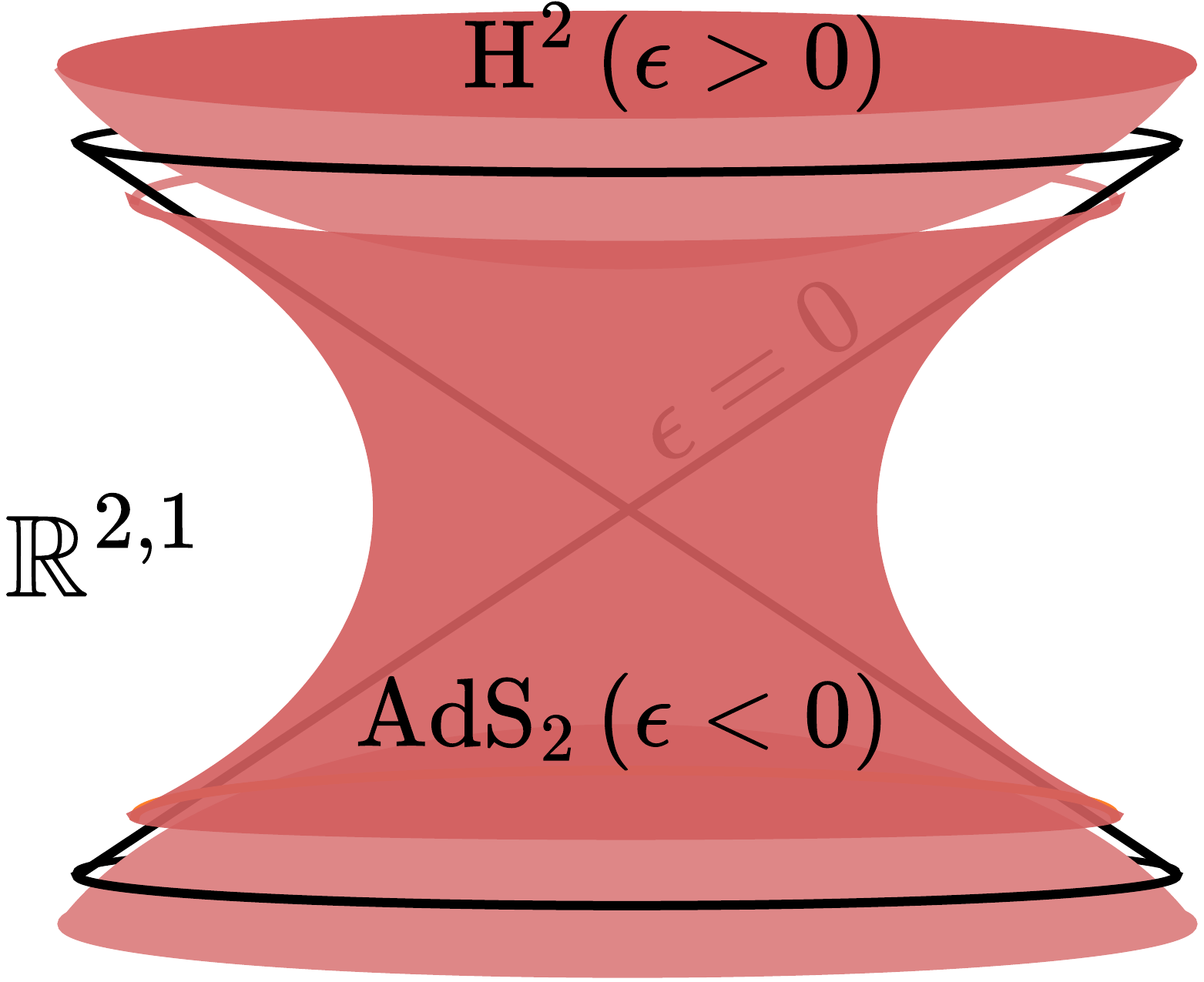}
  \caption{The standard hyperbolic embedding of AdS$_{d+1}$ in $\mathbb R^{2,d}$ foliates one side of the $\mathbb R^{2,d}$'s lightcone ($\epsilon = 0$) with AdS$_{d+1}$ (indexed by $\epsilon < 0$) and Euclidean H$^d$ on the other side (indexed by $\epsilon > 0$). The case for AdS$_2$ in $\mathbb R^{2,1}$ is sketched above. The one exception occurs at $d=2$ when $\ads$ foliates $\rtt$ on \emph{both} sides of its lightcone with $\ads$s with opposite signature (indexed by $\epsilon >0$ and $\epsilon < 0$).}
  \label{fig:R22}
\end{figure}

AdS$_{d+1}$ can be defined in $d$ dimensions through an embedding in $\mathbb R^{2,d}$ as a hyperbolic embedding in terms of its time, $u$ and $v$, and space coordinates, $\{x_i\}_{i=1}^d$:
\begin{equation}
  \label{eq:hyperbolicembedding}
  -u^2 - v^2 + \sum_i^d x_i^2 = \epsilon.
\end{equation}
For $\epsilon <0$ these define Lorentizian hyperbolas (i.e.~AdS$_{d+1}$) with Ricci scalar $R = -6/\epsilon$ and for $\epsilon >0$ they define Euclidean hyperbolas (i.e. H$^d$, see figure~\ref{fig:R22}), both with negative curvature.

An exception occurs for $d=2$ due to the extra split-orthogonal symmetry of $\mathbb R^{2,2}$. For $d=2$ both $\epsilon > 0$ and $\epsilon < 0$ index $\ads$s with Ricci scalar $R = -6/\epsilon$. Both have negative curvature (i.e. $R<0$ for $\det g < 0$ and $R>0$ for $\det g > 0$ for $g$ the metric).

The restriction given by eq.~\ref{eq:hyperbolicembedding} is satisfied by the global coordinates $t$, $\rho$ and $\phi$ defined as,
\begin{align}
& u = \ell \cosh\rho \cos t\\
& v = \ell \cosh\rho \sin t\\
& x = \ell \sinh\rho \cos \phi \\
& y = \ell \sinh \rho \sin\phi,
\end{align}
where $\ell = |\sqrt{\epsilon}|$, and substituting them into the $\rtt$ metric $\text d s^2 = -\text d u^2 - \text d v^2 + \text d x^2 + \text d y^2$ with fixed $\epsilon$ produces the $\ads$ metric:
\begin{equation}
  \text d s^2 = \epsilon(-\cosh(\rho)^2 \text d t^2 + \text d \rho^2 + \sinh(\rho)^2 \text d \phi^2).
\end{equation}

\begin{figure}
  \begin{center}
    \includegraphics[width=0.5\textwidth]{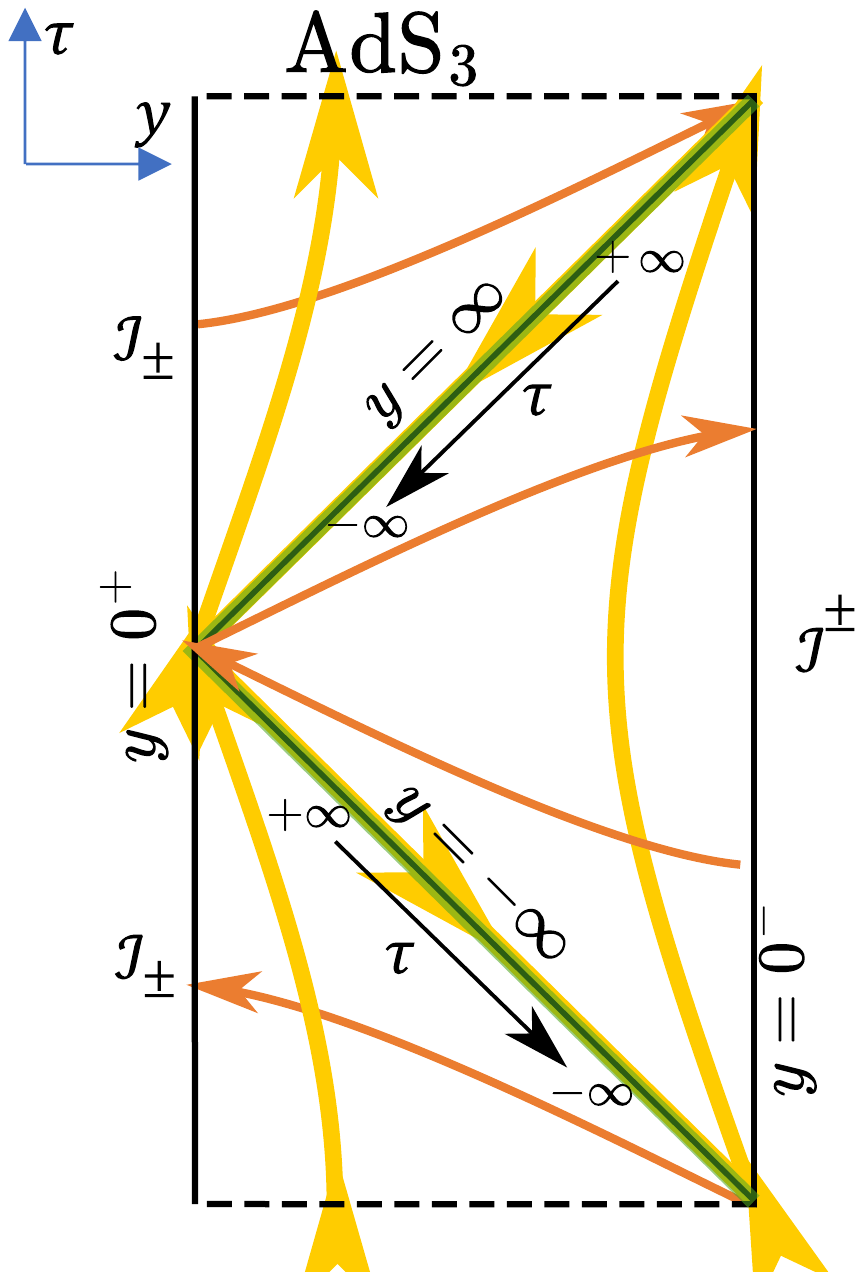}
    \caption{Two Poincar\'e patches of $\ads$. Lightlike periodic geodesics (yellow arrows) discontinuously jump at the Killing horizons (green) at  $y=\pm \infty$ from $\tau = -\infty$ to $\tau = \infty$~\cite{Fujisawa14}. A subset of spacelike geodesics (orange) are also shown, which travel from the $y=0^\pm$ boundary and asymptote at the Killing horizons.}
    \label{fig:Poincare}
  \end{center}
\end{figure}

A very simple coordinate system for $\ads$ is offered by the Poincar\'e coordinates $\{\tau, y, \phi\}$, which give the metric for one-half of $\ads$:
\begin{equation}
  \text d s^2 = \frac{\epsilon}{y^2} (-\text d \tau^2 + \text d x^2 + \text d y^2),
\end{equation}
where $\tau$, $x \in \mathbb R$ and $y \in (0^+, \infty)$. If $y \in (-\infty, 0^-)$ is chosen to coordinatize the other patch with the same set of coordinates consistently, then the two Penrose diagrams of $\ads$ can be sketched as in figure~\ref{fig:Poincare}.

The isometry group of $\rtt$ is $\text{SO}^+(2,2) \sim (\slr \times \slr)/\mathbb Z_2$. The $\slr \times \slr$ part can be seen by identifying the usual Lorentz generators of the spacetime: two rotations ($u \text d v - v \text d u$ and $x \text d y - y \text d x$) and four boosts ($u \text d x + x \text d u$, $u \text d y + y \text d u$, $v \text d x + x \text d v$, and $v \text d y + y \text d v$). Projecting these onto the embedded $\ads$ and recombining produces the following isometry generators:
\begin{align}
  \label{eq:isometriesglobcoords1}
  \chi^\pm_{-1} &= \frac{\ell}{2}\left( \tanh \rho e^{-i (t \pm \phi)}\partial_t + \coth \rho e^{-i(t\pm\phi)} \partial_\phi\right.\\
  & \qquad \left. + i e^{-i(t\pm\phi)} \partial_\rho \right), \nonumber\\
  \label{eq:isometriesglobcoords2}
  \chi^\pm_0 &= \frac{\ell}{2} \left( \partial_t \pm \partial_\phi \right),\\
  \label{eq:isometriesglobcoords3}
  \chi^\pm_{+1} &= \frac{\ell}{2}\left( \tanh \rho e^{i (t \pm \phi)}\partial_t + \coth \rho e^{i(t\pm\phi)} \partial_\phi - i e^{i(t\pm\phi)} \partial_\rho \right).
\end{align}

The $\{\chi^+_i\}_{i\in \{0,\pm 1\}}$ and $\{\chi^-_i\}_{i \in \{0, \pm 1\}}$ subsets produce independent (commuting) $\slr$ isometry groups corresponding to left-handed ($t - \phi$) or right-handed ($t+\phi$) isometries, which means that they possess opposite ``chirality''. In general, the isometries of orientable surfaces do not decompose so neatly into factors of oppositely chiral transformations. To produce a non-orientable manifold, a quotient by a $\mathbb Z_2$ subgroup that is orientation-reversing needs to be taken. 

Note that chirality and orientation are generally unrelated properties. However, as we shall see, embedding a non-orientable manifold in a higher-dimensional manifold can impart a chirality to the manifold from the perspective of the higher-dimensional manifold. Nevertheless, chirality is not a necessary property of a non-orientable manifold.

The $\mathbb Z_2$ factor in $\text{SO}^+(2,2) \sim (\slr \times \slr)/\mathbb Z_2$ has support on both the left- and right-handed $\slr$ isometries. The easiest way to see this is to consider $\rtt$ through its isometry to $\text{Mat}_{2\times2}(\mathbb{R})$:
\begin{align}
  \label{eq:matrixrepM}
 &x^\mu  \leftrightarrow M(x^\mu) = \frac{1}{\sqrt{2}}\begin{pmatrix} u+x & v+y  \\ v-y & x-u \end{pmatrix} \equiv \begin{pmatrix} a& b\\ c& d\end{pmatrix},
\end{align}
where
\begin{equation}
\text{det}[M(\text d x^\mu)] = \eta_{\mu \nu} \text d x^\mu \text d x^\nu = -\text d u^2 - \text d v^2 + \text d x^2 + \text d y^2  
\end{equation}
is the metric and $\text{det}[M(x^\mu)] = -u^2 - v^2 + x^2 + y^2$ is the hyperbolic restriction. Fixing $\det[M(x^\mu)] = \epsilon$ produces $\ads$ and for $\epsilon \equiv 1$ this becomes a matrix in $\slr$ which is closed under matrix multiplication on the left and right, but is unable to distinguish when both are $-I$ matrices and when both are $I$. Therefore, this represents a $(\slr \times \slr)/\mathbb Z_2$-invariance or isometry. 

A larger group containing $\text{SO}(2,2)$ must be considered to obtain discrete symmetries that break orientation. In particular, orbifolding $\ads$ by orientation-changing $\mathbb Z_2$ identifications that are part of $\text{O}(2,2)$ but not $\text{SO}(2,2)$ produces ``orientifolds'' of $\ads$~\cite{Loran10}.

$\text{SO}^+(2,2)$ has fundamental group $\mathbb Z \times \mathbb Z$, which has three subgroups of index two. This implies that there exist three corresponding double covers: the double cover $\text{Spin}^+(2,2)$ corresponds to the subgroup $\{I\times I, -I \times -I\}$, and the two others, which we will denote $\text{Spin}^+_L(2,2)$ and $\text{Spin}^+_R(2,2)$, correspond to the subgroups $\{I \times I, \pm I \times I\}$ and $\{I \times I, I \times \pm I\}$, respectively. $\text{Spin}^+(2,2)$ is a ``diagonal'' spinorial double cover of $\text{SO}^+(2,2)$ , which means that it can be considered to evenly double cover the two $\slr$ factors in $\text{SO}^+(2,2)$. $\text{Spin}_{L/R}(2,2)$ are ``off-diagonal'' double covers because they double cover opposite factors of $\slr$. Specifically, $\text{Spin}^+_L(2,2) \equiv (\mplr \times \slr)/\mathbb Z_2$ and $\text{Spin}^+_R(2,2) \equiv (\slr \times \mplr)/\mathbb Z_2$, where $\mplr$ (the ``metaplectic'' group) is the double cover of $\slr$ whose $\mathbb Z_2$ central extension commutes with the overall $\mathbb Z_2$ quotient in the definitions of $\text{Spin}^+_{L/R}(2,2)$. 

The off-diagonal $\text{Spin}^+_{L/R}(2,2)$ double covers are more relevant to us than the diagonal $\text{Spin}^+(2,2)$ because breaking orientation requires introducing a two-dimension Mobius strip into the spacetime. When this Mobius strip is embedded in a higher-dimensional space such as $\rtt$, it can twist clockwise or counter-clockwise, i.e.~it inherits a chirality, which can be aligned to $(t\pm \phi)$. We will show that choosing whether the intermediate double cover is $\text{Spin}^+_L(2,2)$ or $\text{Spin}^+_R(2,2)$, will effect whether the resultant non-orientable spacetime is represented clockwise or counter-clockwise, and the two different representations will be related by a mirror symmetry. This will greatly simplify its CS theory, which will exhibit this mirror symmetry between its two otherwise independent chiral actions.

$\ads$ projectively inherits its properties from the simpler $\rtt$ embedding spacetime. This will allow us to introduce new non-trivial boundary conditions by putting them in as a bulk orientifold fixed point in the double cover of $\rtt$ in a region that is disjoint from the embedded $\ads$. As long as the two regions asymptotically approach each other, the $\ads$ will also projectively inherit the new topology on its boundary~\cite{Loran10}.

In the next subsection, we will derive a new non-orientable $\rtt$ spacetime, which will be an orientifold of its double cover, $\text{Spin}^+(2,2)$, and we will particularly make use of the off-diagonal $\text{Spin}^+_{L/R}(2,2)$ double covers. 
We will then orientifold this double cover by introducing a topological twist in its bulk.

\subsection{Derivation of non-orientable $\rtt$}
\label{sec:doublecover}

The two equivalent $\ads$ embedded in $\rtt$ can be mapped to each other by an exchange of space-like and time-like coordinates, for example an exchange operation $\sigma_Z$ defined as:
\begin{equation}
  \label{eq:Mobiustwist}
\sigma_Z [(u, v, x, y)] = -(x, y, u, v).
\end{equation}
Our objective is to leverage this observation, and to create an embedding spacetime and a resultant $\ads$ spacetime embedded in it which also includes the space-time coordinate exchange above as a symmetry.

To do so, we go into four-dimensional null coordinates defined as $(a, b, c, d) = (x+u, v+y, x-u, v-y)$.
The line element becomes
\begin{equation}
ds^2 = \dee a \dee d - \dee b \dee c, \ (a, b, c, d) \in \mathbb{R}^4.
\end{equation}
As a result, the space-time coordinate exchange above becomes $\sigma_Z \circ(a, b, c, d) \leftrightarrow (-a, -b, c, d)$.
Working in null coordinates simplifies the analysis as we only need to keep track of the $(a, b)$ coordinates.
We note also that the space-time reflection operation is a reflection in four dimensional null coordinates about the reflection plane $(0, 0, c, d)$, which is stabilized by the exchange operation.

A first attempt would be to quotient $\rtt$ by the exchange operation.
The issue is that the ordinary $\rtt$ metric is not invariant under $\sigma_Z$, as it is not an isometry of the spacetime. 
To fix this, we consider an altered metric over the domain $\mathbb{R}^4$ which is invariant under $\sigma_Z$:
\begin{equation}
  \dee s_{\widetilde{\rtt}}^2 = \dee (a^2 - b^2) \dee d - \dee (2ab) \dee c, \ (a, b, c, d) \in \mathbb{R}^4.
  \label{eq:r22doublecovermetric}
\end{equation}
The particular form of the metric is a result of a ``realification'' of the complex square function:
\begin{align}
z \equiv a + ib \rightarrow z^2 =a^2 - b^2 + i(2ab).
\end{align}
This choice was made because the complex square is naturally invariant under $(a, b) \rightarrow (-a, -b)$, equivalently $z \rightarrow -z$, and also has the minimal singular behavior of a function with this property: $z^2$ is 2:1 over the complex plane except at $z = 0 \rightarrow a = b = 0$, which is precisely the stabilized reflection plane of the operation $\sigma_Z$. Eq.~\ref{eq:r22doublecovermetric} actually possesses a double cover of $\rtt$'s isometry group, $\text{SO}^+(2,2)$, which can be identified as $\text{Spin}^+_R(2,2)$ (see appendix~\ref{sec:spindoublecover} for more details).

The final step is to take the quotient $(a, b, c, d) \sim \sigma_Z\circ (a, b, c, d)$. 
Since the metric is invariant under this operation, it carries through, but the domain in the $(a, b)$ plane must be restricted under the quotient $(a, b) \sim (-a, -b)$.
The resulting domain is:
\begin{align}
\frac{\mathbb{R}^2}{\mathbb{Z}_2} \equiv \left\{ [a, b]: (a, b) \sim (-a, -b), \ (a, b) \in \mathbb{R}^2 \right\}.
\end{align}
Similar to the antipodal quotient of a 2-sphere producing $\mathbb{RP}^2$, we can understand $\mathbb{R}^2/\mathbb{Z}_2$ as a half space with a special boundary line that is antipodally identified with itself:
\begin{align}
\frac{\mathbb{R}^2}{\mathbb{Z}_2} \cong \{ (a, b) \in \mathbb{R}^2 : a > 0 \} \cup \{ (a, b) \in \mathbb{R}^2 : a = 0, \ b \sim -b \}.
\end{align}
With this quotient in place, we define our new embedding space geometry:
\begin{align}
  ds^2_{\widetilde{\rtt}/\mathbb Z_2} &= \dee (a^2 - b^2) \dee d - \dee (2ab) \dee c,\\
  &(a, b) \in \frac{\mathbb{R}^2}{\mathbb{Z}_2}, \ (c, d) \in \mathbb{R}^2.\nonumber
\label{eq:quotient-geometry}
\end{align}
We will refer to this spacetime as $\widetilde{\rtt}/\mathbb Z_2$, where $\widetilde{\rtt}$ is the double cover of $\rtt$.

Taking the $\mathbb Z_2$ quotient (i.e.~halving the $(a,b)$ domain) produces an orientifold of $\rtt$ because $\sigma_Z \in \mathbb Z_2$ is metric-sign-reversing in terms of the original $\rtt$ coordinates (and thus orientation-reversing): $\dee a \dee d - \dee b \dee c = \text d s^2_{\rtt} \underset{\sigma_Z}{\rightarrow} - \dee s^2_{\rtt}$. In the following, we start to characterize the consequences of this quotient on the single cover of the space.

\subsubsection{Tangent bundle}
Consider the following tentative coordinate transformation:
\begin{align}
&\Phi: \frac{\mathbb{R}^2}{\mathbb{Z}_2} \times \mathbb{R}^2 \rightarrow \mathbb{R}^{2,2} \\
&(a, b, c, d) \rightarrow (a^2-b^2, 2ab, c, d).
\end{align}
Defining $z \equiv a+ib$ as before and $w \equiv c+id$, this can be also represented over the complex numbers as:
\begin{align}
&\Phi: \frac{\mathbb{C}}{\mathbb{Z}_2} \times \mathbb{C} \rightarrow \mathbb{C}^2 \\
& (z, w) \rightarrow (z^2, w).
\end{align}
We note that $\Phi$ is a bijective function and is differentiable over its domain. However, the inverse is not differentiable nor continuous everywhere on its domain:
\begin{align}
&\Phi^{-1}: \mathbb{C}^2 \rightarrow  \frac{\mathbb{C}}{\mathbb{Z}_2} \times \mathbb{C} \\
& (z, w) \rightarrow (z^{1/2}, w).
\end{align}
This is because $z^{1/2}$ has a discontinuity at the branch cut at the negative reals in $\mathbb{C}$.
As such, this is not a true coordinate transformation on the entire domain. However, we can consider $\Phi$ as a valid coordinate choice on a patch which excludes the boundary of the half-space at $a = 0$
\begin{align}
& U_1 \equiv \left(\frac{\mathbb{R}^2}{\mathbb{Z}_2} \times \mathbb{R}^2\right) \setminus \{a= 0\}.
\end{align}

Defining the single cover of this patch in coordinates $(A, B) \equiv (a^2-b^2, 2ab)$, we find that the line element over the patch becomes:
\begin{align}
d s^2\bigr|_{U_1} = \dee A \dee d - \dee B \dee c, \ (A, B, c, d) \in \mathbb{R}^4.
\end{align}
Abstractly, we can thereby write:
\begin{align}
\widetilde{\rtt}/\mathbb Z_2 \setminus \{a = 0\} \cong \rtt.
\end{align}
However, $\widetilde{\rtt}/\mathbb Z_2$ is not itself diffeomorphic to $\rtt$, and we want to characterize how it is different.

To do so, we can look at how different geometrically relevant quantities transform under the coordinate transformation above.
Of particular interest are the tangent space basis vectors and co-tangent space basis 1-forms.
These transform as:
\begin{align}
& \begin{pmatrix} \partial_A \\ \partial_B \end{pmatrix} = \frac{1}{2\sqrt{R}} \begin{pmatrix} \cos(\Theta/2) & -\sin(\Theta/2) \\ \sin(\Theta/2) & \cos(\Theta/2) \end{pmatrix} \cdot \begin{pmatrix} \partial_a \\ \partial_b \end{pmatrix} \\
& \begin{pmatrix} \dee A \\ \dee B  \end{pmatrix} = \sqrt{R} \begin{pmatrix} \cos(\Theta/2) & -\sin(\Theta/2) \\ \sin(\Theta/2) & \cos(\Theta/2) \end{pmatrix} \cdot \begin{pmatrix} \dee a \\ \dee b \end{pmatrix},
\end{align}
where we use polar coordinates $(A,B) =R(\cos\Theta, \sin\Theta)$.
A proof for these relations can be found in appendix~\ref{sec:tangent-transformations}.

Due to the factor of $1/2$ in the angular arguments above, we find that the tangent bases and co-tangent bases are double valued on the manifold, namely as we take $\Theta \rightarrow \Theta + 2\pi$ which takes you back to the same point on the manifold, the tangent vectors will go from $(\partial_A, \partial_B, \partial_C, \partial_D) \rightarrow (-\partial_A, -\partial_B, \partial_C, \partial_D)$.
This is not what happens in ordinary $\rtt$, and indicates a lack of orientability in the spacetime analogous to what occurs on a Mobius strip. 
In the Mobius case, a single rotation around the base $\mathbb{S}^1$ will reflect the tangent vector basis along the fiber direction. 
This is our first indication that there is no way to consistently pick an orientation on the tangent and co-tangent bundles. 
A more explicit demonstration of the non-orientability follows in the next section.

\subsubsection{Mobius topology of the light cone}
\label{sec:mobius}
Non-orientability of a spacetime becomes especially apparent when a Mobius strip topology is identified in the space. 
To do so, we will compare at the light cone (relative to a fixed origin) between ordinary $\rtt$ and the quotient space $\widetilde{\rtt}/\mathbb Z_2$:
\begin{align}
& \rtt \rightarrow ad - bc = 0\\
& \widetilde{\rtt}/\mathbb Z_2 \rightarrow (a^2-b^2)d - (2ab)c = 0.
\end{align}
Taking polar coordinates $(a, b, c, d) = (R\cos\Omega, R\sin\Omega, r\cos\omega, r\sin\omega)$, we get the expressions:
\begin{align}
  \label{eq:rttpolar}
  & \rtt \rightarrow Rr \sin(\Omega - \omega) = 0 \\
  \label{eq:rttzpolar}
  & \widetilde{\rtt}/\mathbb Z_2 \rightarrow R^2r \sin(2\Omega - \omega) = 0.
\end{align}

\begin{figure}[htbp]
\begin{center}
  \includegraphics[width=0.5\textwidth]{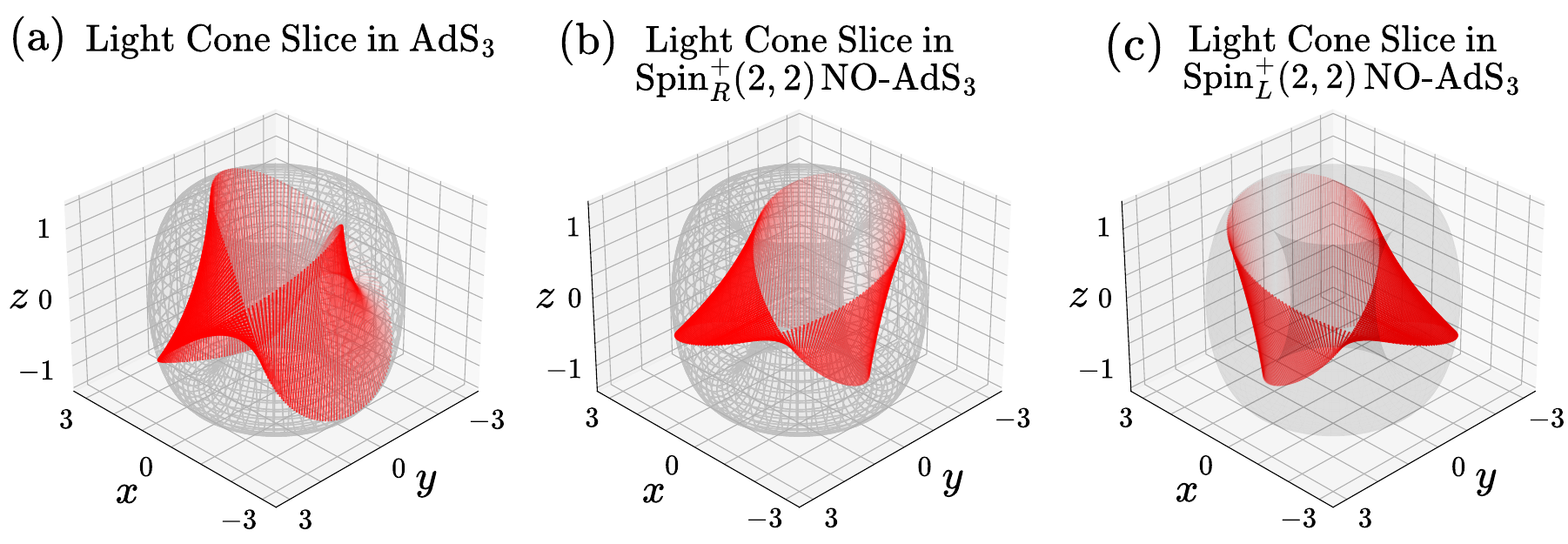}
  \caption{
    Consider a slice at fixed $R = 1$ for eq.~\ref{eq:rttpolar}-\ref{eq:rttzpolar}. The remaining coordinates $\Omega \in \mathbb{S}^1$ and $(r, \omega) \in \mathbb{R}^2$ can be plotted as a (red) slice of the solid torus embedded in $\mathbb R^3$ with coordinates $x$, $y$ and $z$. The slice's $(r, \omega)$ parameterizes the disk at a given azimuthal angle $\Omega$. 
    (a) For $\rtt$, the result is a manifold consisting of two Mobius strips glued end-to-end, which thereby untwist each other, and are topologically equivalent to a cylinder (the orientable double cover of a Mobius strip).
    (b) On the other hand, for the $\rttz$ constraint equation, the result is a Mobius strip embedded in the torus.
 This is the effect of the factor of two in front of $\Omega$ in the $\rttz$ constraint. (c) Applying the same approach for $\rtt/\mathbb Z_2 \equiv (a^2 - c^2) d - (2ac) b = 0$ can be found to produce a Mobius strip with an opposite chirality.
}
\label{fig:light-cones}
\end{center}
\end{figure}

 A plot of the solutions are shown in figure~\ref{fig:light-cones} and demonstrate how $\widetilde{\rtt}/\mathbb Z_2$ (eq.~\ref{eq:rttzpolar}) possesses a Mobius twist whereas $\rtt$ (eq.~\ref{eq:rttpolar}) possesses a cylindrical topology. Doing the same thing with $\widetilde{\rtt}/\mathbb Z_2 \rightarrow (a^2-c^2)d - (2ac)b=0$ produces a Mobius twist with opposite chirality and corresponds to the other $\text{Spin}^+_L(2,2)$ double cover.

\subsubsection{Isometries}

Local isometries of $\widetilde{\rtt}/\mathbb Z_2$ are the same as for $\rtt$, $\slral \oplus \slral$, as expected for an orbifold. However, the algebra of isometries does reduce if one considers invariance under the global $\mathbb Z_2$ quotient.

$\widetilde{\rtt}/\mathbb Z_2$ has the unique property that under a rotation $\Theta \rightarrow \Theta + 2\pi$ in the $(A, B)$ plane, the tangent vectors are negated in $(\partial_A, \partial_B)$, and so the only Killing vectors which can survive this are those which are invariant under $(A, B, C, D) \sim (-A, -B, C, D)$.
The remaining Killing vectors are broken since they cannot be globally defined as a vector field on $\widetilde{\rtt}/\mathbb Z_2$.

Since $\rttz$ has a line element which is almost everywhere $\rtt$, we start with the (non-translational) Killing vectors of $\rtt$ in $(u, v, x, y)$ coordinates:
\begin{align}
& u\partial_v - v\partial_u,\qquad x\partial_y - y\partial_x \\
& u\partial_y + y\partial_u,\qquad  v\partial_x + x\partial_v \\
& u\partial_x + x\partial_u,\qquad  v\partial_y + y\partial_v.
\end{align}
Taking linear and antilinear combinations produces:
\begin{align}
& J_\pm = u\partial_v - v\partial_u \pm (x\partial_y - y\partial_x) \\
& K_\pm = u\partial_y + y\partial_u \pm (v\partial_x + x\partial_v) \\ 
& H_\pm = u\partial_x + x\partial_u \mp (v\partial_y + y\partial_v).
\end{align}
The quotient above is equivalent to $(u, v, x, y) \sim (-x, -y, -u, -v)$, which leaves only $J_+, K_+, H_\pm$ invariant but takes $J_-, K_- \rightarrow -J_-, -K_-$.
As such, we find that four Killing vectors are preserved, while two are broken.

The commutation rules for the remaining four generators are:
\begin{align}
& [H_-, H_+] = [H_-, K_+] = [H_-, J_+] = 0 \\
& [H_+, K_+] = 2J_+, \ [H_+, J_+] = 2K_+, \  [J_+, K_+] = 2H_+,
\end{align}
which makes the global isometry algebra
\begin{align}
\mathfrak{iso}(\rttz) = \slral \oplus \mathfrak{R},
\end{align}
where the central element $\mathfrak{R}$ is generated by $H_-$.

This is the same as the global isometry group of self-dual orbifolds of $\ads$~\cite{Coussaert98}. We shall later identify the orientifolds of these~\cite{Loran11} as excited states of $\spacetime$.

\subsection{Derivation of non-orientable $\ads$}
\label{sec:poincarepatches}

$\mathbb Z_2$ orbifolds or orientifolds of a spacetime $\mathcal M$ can be considered to be two glued together $\mathcal M$ base spaces, $\mathcal M_1$ and $\mathcal M_2$, which are isometric to the original space. 
As discussed in section~\ref{sec:preamble}, Poincar\'e patches provide a particular simple coordinate system for $\ads$. In the prior subsection we considered a transformation from $\rtt$ to a double cover $\widetilde{\rtt}$ (a two-to-one isometry) that transformed its hyperbolic restrictions $ad-bc$ to $(a^2-b^2)d-2abc$ (and similarly to its line element). Here, we will consider a slightly different transformation from $ad-bc$ to two $\pm a(ad-bc)$ patches instead, because this leads to a particular simple way to view how the orientifold on this double cover produces non-orientable embedded $\ads$s in terms of their Poincar\'e patches.

\subsubsection{Construction as two glued patches}
\label{sec:constructionastwopatches}

Given the four-dimensional embedding spacetime $\rtt$, we define a diffeomorphic coordinate change replacing $a^2-b^2$ by $\pm a'^2$ and $2ab$ by $a'b' \implies b' = \pm\frac{2 a b}{a'^2}$. 
To do this we consider a new isometry between $\mathbb R^{2,2}$ and a matrix $M'(x^\mu) = \left( \begin{array}{cc} a'& b'\\ c'& d'\end{array}\right)$ with $\det[M'(x^\mu)] = a (a d - b c)$. 
Since isometries preserve the Ricci scalar $R$, setting $\det[M'(x^\mu)] = \epsilon$ will necessarily define an AdS$_3$ spacetime on the $a (ad-bc) = \epsilon$ slice with $R=-6/\epsilon$. Given that $M'$'s determinant is a quadratic function of $M$'s elements (see eq.~\ref{eq:matrixrepM}), it follows that $M'$ needs to be a quadratic function of $M$ (i.e.~it will locally be a two-to-one isometry):
\begin{equation}
  \label{eq:twotooneisometry}
  M' = M^T \bar A M + \bar B M,
\end{equation}
where $\bar A = \left(\begin{array}{cc} \bar A_{11}& \bar A_{12}\\ \bar A_{21}& \bar A_{22} \end{array}\right)$ and $\bar B = \left(\begin{array}{cc} \bar B_{11}& \bar B_{12}\\ \bar B_{21}& \bar B_{22} \end{array}\right)$ are in $\slr$ (the transpose of $M$ is necessary in the quadratic factor to preserve the orthonormal isometries of the space).

Computing the determinant reveals that
\begin{align}
  \label{eq:detM'}
  \det M' =& (a d-b c) \left(m'_{bc-ad} (b c-a d) \right.\\
  & \left.+ m'_a a + m'_b b + m'_c c + m'_d d + m'_1\right), \nonumber
\end{align}
where
\begin{align}
  m'_a &= \left(\bar A_{11} \bar B_{22}-\bar A_{12} \bar B_{21}\right), m'_b = \left(\bar A_{12} \bar B_{11}-\bar A_{11} \bar B_{12}\right),\nonumber\\
  m'_c &= \left(\bar A_{21} \bar B_{22}-\bar A_{22} \bar B_{21}\right), m'_d = \left(\bar A_{22} \bar B_{11}-\bar A_{21} \bar B_{12}\right),\\
  m'_1 &= \left(\bar B_{11} \bar B_{22}-\bar B_{12} \bar B_{21}\right), m'_{bc-ad} = \left(\bar A_{12} \bar A_{21}-\bar A_{11} \bar A_{22}\right). \nonumber
\end{align}

Setting $\bar A = \left( \begin{array}{cc} \pm 1& 0\\ 0& 0\end{array}\right)$ and $\bar B = \left( \begin{array}{cc} 0& 0\\ 0& 1\end{array}\right)$ produces our desired isometry between $\mathbb R^{2,2}$ and $M'$ where $\det M' = \pm a (ad - bc)$, which we will shortly identify with the $\text{Spin}^+_R(2,2)$ double cover of $\mathbb R^{2,2}$, $\widetilde{\mathbb R}^{2,2}$:
\begin{align}
  \widetilde{\mathbb R}^{2,2} \equiv M'(a,b,c,d) &= \begin{pmatrix} \pm a^2 & \pm a b \\ \pm a b + c & \pm b^2 + d \end{pmatrix},
\label{eq:invariant_map}
\end{align}
for $a,b,c,d \in \mathbb R$.

As mentioned, maintaining the isometry between $M$ and $M'$ means that setting $\det M' = \pm \epsilon$ for $\epsilon > 0$ corresponds to an embedded manifold with Ricci scalar $R = -\frac{6}{\epsilon}$, which is (the double cover of) an $\ads$ spacetime. 

We now take stock of what we have done. The first term in eq.~\ref{eq:twotooneisometry} produces a double cover of $a$ and $b$, which is added to the second term's single cover of $c$ and $d$. Since right-hand $\slr$ isometries of $M$ act by right matrix multiplication on $M$, this means that they act on $a$ and $b$ separately from $c$ and $d$. As a result, the right-hand $\slr$ isometry factor is doubly covered by eq.~\ref{eq:twotooneisometry}'s quadratic map. On the other hand, the left-hand $\slr$ isometries act on $a$ and $c$ separately from $b$ and $d$, which means that they are unaffected by eq.~\ref{eq:twotooneisometry}'s map. 
As a result, this describes a $\text{Spin}^+_{R}(2,2)$ double cover of $\text{SO}^+(2,2)$. 

We are now interested in orientifolding this spacetime with the $\mathbb Z_2$ quotient corresponding to the new symmetry, $(a,b) \sim (-a,-b)$. To do this, we absorb the squared $a^2$ coordinate into a new coordinate $1/y$ such that $y>0$ corresponds to the $+\epsilon$ case and $y<0$ corresponds to the $-\epsilon$ case. We also introduce the coordinates $\tau$ and $x$ through ${(\tau - x)}/{y} = ab$ and ${2x}/{y} = c$, so that the matrix representation can be rewritten as
\begin{equation}
  \spacetime
  \coloneqq \begin{cases}
    \frac{1}{y} \begin{pmatrix}  1 & \tau - x \\ \tau + x &  \tau^2 - x^2 - \epsilon y^2 \end{pmatrix} & \text{for } y \in (0,\infty),\\
    \frac{1}{y} \begin{pmatrix}  1 & x - \tau \\ x + \tau &  x^2 - \tau^2 - \epsilon y^2 \end{pmatrix} & \text{for } y \in (-\infty,0).
    \end{cases}
\label{eq:space}
\end{equation}

Note that unlike the $M(x^\mu)$ associated with regular $\ads$, these matrices do not form a group under matrix multiplication. 

The reason for picking these coordinates becomes apparent upon computing its line element from $\text{det}[M'(\text d x^\mu) ]$ with respect to $\tau$, $x$, $y$:
\begin{align}
  \text{d}s^2 %
  &= y^{-2}(-\text{d} \tau^2 + \text{d}x^2 + \epsilon \text{d}y^2),
    \label{eq:Poincarepatch1}
\end{align}
where $\tau$, $x \in \mathbb{R}$, and $y \in (0^+, \infty)$ and
\begin{align}
  \text{d}s^2 %
  &= y^{-2}(\text{d}\tau^2 -\text{d}x^2 + \epsilon \text{d}y^2),
    \label{eq:Poincarepatch2}
\end{align}
where $\tau$, $x \in \mathbb{R}$, and $y \in (-\infty, 0^-)$. These are two Poincar\'e patches glued together at $y=\pm \infty$. As we showed in section~\ref{sec:preamble}, $\ads$ can be identically split into two Poincar\'e patches except that the patches here have exchanged $\tau$ and $x$ coordinates. This means they are twisted with respect to each other (the coordinates in one patch cannot be solely exchanged back without breaking their global consistency with the second patch). 
Like in the orientable case, the asymptotic boundary of $\ads$ of these two Poincar\'e patches is at $y = 0$ and there is a Killing horizon at $y=\pm \infty$ where they are glued together. 

We can see the effect of a Mobius twist between these two patches by examining how the spacetime is orbifolded. Whereas a $\mathbb Z_2$ quotient of $\ads$ can be described as two $\ads$ glued together, the $\mathbb Z_2$ quotient of its double cover corresponds to two \emph{halves} of $\ads$ glued together. After our diffeomorphism from $a$ and $b$ coordinates to $a'$ and $b'$, we have found that these two halves of $\ads$ correspond to Poincar\'e patches of $\ads$. It is known that at the conformal boundary, $\tau$ and $x$ become collinear with the global boundary coordinates $t$ and $\phi$, respectively (i.e.~this is a static patch of an observer at $y\rightarrow 0$). Since these are exchanged in the two Poincar\'e patches given by eqs.~\ref{eq:Poincarepatch1}-\ref{eq:Poincarepatch2}, and $(t+\phi)$ and $(t-\phi)$ parameterize the two factors of $\slral$ in the original isometry algebra of $\ads$ (see eqs.~\ref{eq:isometriesglobcoords1}-\ref{eq:isometriesglobcoords3}), we see that asymptotically the $\mathbb Z_2$ quotient imposes a mirror symmetry between the two $\slral$ factors. 

\begin{figure}[ht]
\begin{center}
\includegraphics[width=0.5\textwidth]{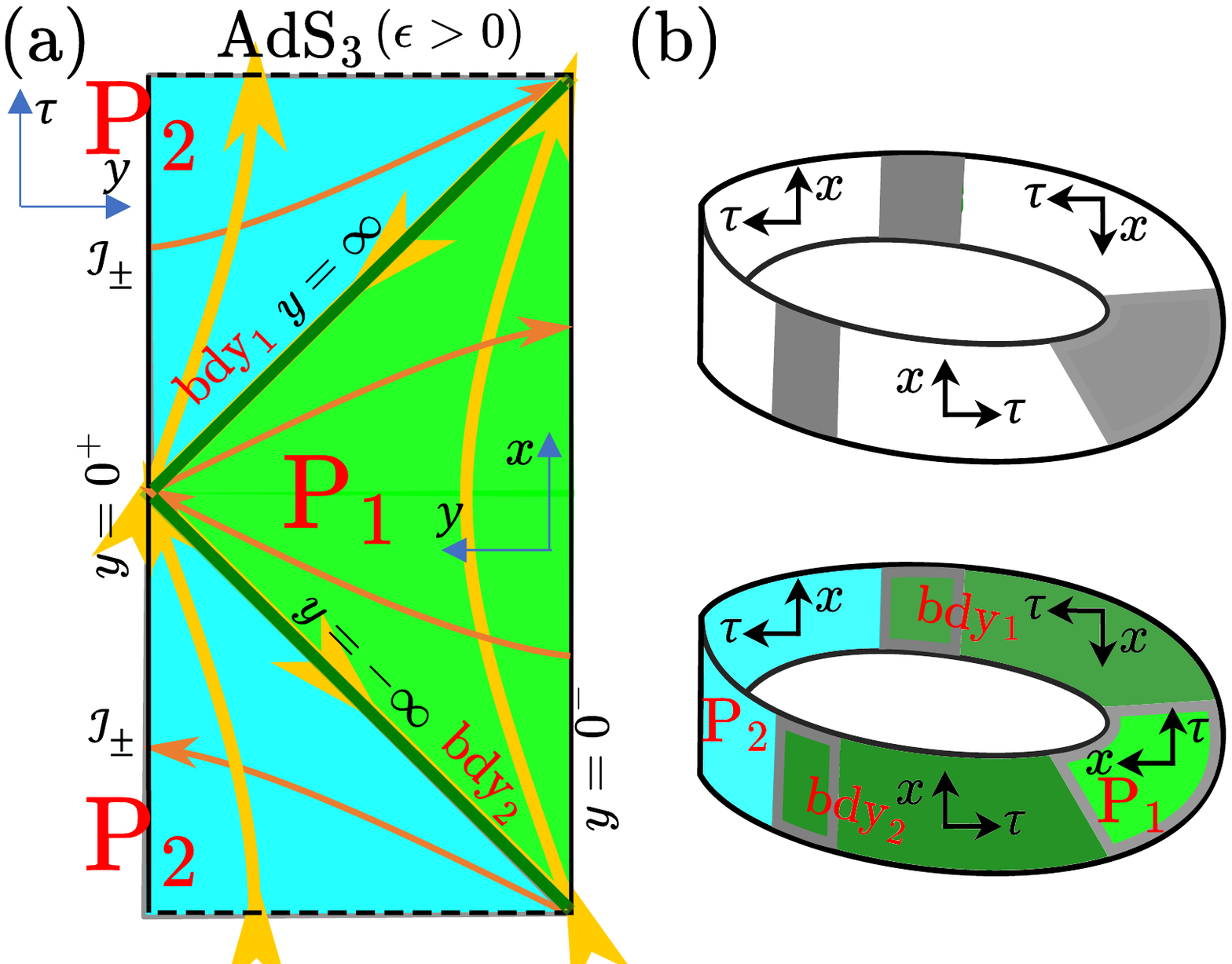}
\caption{(a) Two Poincar\'e patches of $\spacetime$ denoted P1 and P2. Show are some characteristic spacelike geodesics (orange arrows) and closed timelike geodesics (yellow arrows), which traverse the Killing horizon boundaries (denoted `bdy1' for $\tau,-x>0$ and `bdy2' for $\tau,-x<0$). (b) Mobius strip defined in terms of three overlapping patches with defining axes shown. Their overlaps are shaded in grey. The first Poincar\'e patch (P1) corresponds to the first Mobius patch (shaded cyan) and the second Poincar\'e patch (P2) corresponds to the overlap between the remaining two Mobius patches (shaded light green). The boundaries between P1 and P2 correspond to the remaining Poincar\'e patches and overlaps (shaded dark green).}
\label{fig:Mobius}
\end{center}
\end{figure}

Timelike geodesics are periodic and half of their period lies in one Poincar\'e patch and the other half in the other Poincar\'e patch, which means that they traverse $y=\pm \infty$ twice. Since the Killing horizon at $y=\pm \infty$ is not part of the Poincar\'e patches, technically at least one more overlapping patch is necessary to describe the full spacetime with two Poincar\'e patches. 
The same is true for the two-dimensional Mobius strip~\cite{Chen14}. To define it requires at least three overlapping charts, where each chart must have one of its two axes oriented oppositely to one of the axes of each adjacent chart. As a result, as shown in figure~\ref{fig:Mobius}, the two Poincar\'e patches above can be associated with one patch of the Mobius strip and the overlapping region between the remaining two Mobius patches.

This means that when stationary timelike observers following closed timelike geodesics go from one Poincar\'e patch to another and then back again, they will return with their tangent vectors pointing in the opposite orientation compared to how they began, and they will need to double their orbit to come back with the same orientation. 
This also explains the asymptotic mirror symmetry between the two $\slral$ factors; when embedded in more than two dimension, counter and counter-clockwise twisting Mobius strips are related by a mirror symmetry.

When $y \rightarrow 0$, the two patches should approach the same boundary. However, in this limit it appears that the metric diverges with opposite signatures:
\begin{equation}
  \text d s^2 = \lim_{y \rightarrow 0} \pm (-\text d \tau^2 + \text d x^2)/y^2.
\end{equation}
This can only remain consistent if the metric degenerates into a null submanifold at the boundary, such that $\tau$ and $x$ become null directions. As mentioned earlier, at $y \rightarrow 0$, $\tau$ and $x$ become collinear to global $t$ and $\phi$. Hence, the boundary of non-orientable $\ads$ must consist of two null dimensions compared to regular $\ads$'s timelike and spacelike directions.

Just like in regular $\ads$, unrolling the periodic time dimension of this space will get rid of closed timelike curves. However, unlike the regular case, the remaining compact periodic dimension at the boundary will now be null-like (lightlike) instead of spacelike. The same property is observed in ``self-dual'' orbifolds of $\ads$, where the $\mathbb Z$ quotient acts on only one $\slr$ factor~\cite{Coussaert98,Witten07}. Again, we will see later (in section~\ref{sec:moreexactCS}) that the orientifolds of these states~\cite{Loran11} become the natural excited states of this non-orientable $\ads$ spacetime. 

Finally, we note that the Ricci scalar is continuous everywhere on this non-orientable $\ads$, including at all points asymptotically on its boundary. Similar behavior is observed for $\ads$ orientifolds for lightlike orientifold fixed surfaces, where the geometry was found to have a continuous Ricci scalar at the fixed surface~\cite{Loran11}. In both cases, this means that the spacetimes are exact solutions to Einstein's equations everywhere, i.e.~they do not correspond to conical singularity spacetimes.

\subsubsection{Isometries}
\label{sec:adsisometries}

For a local patch which excludes the ``twist'', the metric of $\spacetime$ is the same as ordinary $\ads$, again with the non-trivial rotation of the tangent vectors.
Since it is an isometric embedding, we infer that the local isometries are the same, which is to say they are $\slral \oplus \slral$. On the other hand, we expect the global isometries to be equivalent to $\widetilde{\rtt}/\mathbb Z_2$, which is $\slral \oplus \mathfrak{R}$. However, the asymmetric fixed point of the quotient in $\mathbb R_{2,2}$ with respect to the embedded $\ads$s should break this global algebra further for the double cover of $\spacetime$ at its conformal boundary.

To understand what occurs at the conformal boundary, we will look at the Killing vectors in $\ads$ global coordinates as $\rho \rightarrow \infty$.
The Killing vectors are given by eqs.~\ref{eq:isometriesglobcoords1}-\ref{eq:isometriesglobcoords3}. 
The coordinate exchange $(u, v, x, y) \rightarrow (-x, -y, -u, -v)$ is then equivalent to:
\begin{align}
 \ell &\rightarrow i \ell,\\
 \rho &\rightarrow \arcsinh( -i \cosh(\rho)),\\
 (t, \phi) &\rightarrow (\phi + \pi, t + \pi).
\end{align}
Let us examine this equivalence relation from the perspective of the double cover $\text{Spin}^+_R(2,2) = (\slr \times \mplr)/\mathbb Z_2$. 

Asymptotically as $\rho\rightarrow \infty$, the Killing vectors reduce to:
\begin{align}
& \chi^\pm_{-1} = e^{-i(t\pm\phi)}\partial_t \pm e^{-i(t\pm\phi)}\partial_\phi\\
& \chi^\pm_{0} = \partial_t \pm \partial_\phi\\
& \chi^\pm_{1} = e^{i(t\pm\phi)}\partial_t \pm e^{i(t\pm\phi)}\partial_\phi,
\end{align}
with the coordinate exchange reduced to just the angular part above.
Under the coordinate exchange we find that the $\chi^+$ Killing vectors are unchanged, as $t + \phi \rightarrow t+ \phi$, and are hence preserved.
The vector $\chi^-_0 \rightarrow -\chi^-_0$, and is certainly broken.
The remaining two have the properties $\chi^-_{-1} \rightarrow -\chi^-_1$ and $\chi^-_1 \rightarrow -\chi^-_{-1}$.
We find then that the anti-linear combination is preserved.
Defining $\omega_\pm \equiv t \pm \phi$, we can write the four remaining Killing vectors as:
\begin{align}
& \chi^+_0 = \partial_{\omega_+} \\
& \chi^+_1  = \left(\cos(\omega_+) + \sin(\omega_+)\right)\partial_{\omega_+}\\
& \chi^+_{-1} = \left(\cos(\omega_+) - \sin(\omega_+)\right)\partial_{\omega_+} \\
& \chi^-_1 - \chi^-_{-1} = \sin(\omega_-)\partial_{\omega_-},
\end{align}
where
\begin{equation}
  \sigma_z (\omega_{\pm}) = \pm \omega_{\pm}.
\end{equation}
Note, however, that $\sin(\omega_-)\partial_{\omega_-}$ is not non-vanishing everywhere, and hence cannot be a Killing vector field at the conformal boundary.
This vanishing feature is not found in the bulk of the $\spacetime$ spacetime, since there one retains the dependence on $\partial_\rho$.
Hence, we find that the isometries asymptotically reduce to only $\slral$, and so the global isometry algebra is $\mathfrak{iso}(\lim_{\rho \rightarrow \infty}\spacetime) = \slral$.

This demonstrates that at the fixed point of the orientifold, $\spacetime$ only possesses one local $\slral$ isometry. This is a result of the $\mathbb Z_2$ quotient's introduction of mirror symmetry between the chiral and anti-chiral $\slr$ factors of regular $\ads$. This property will be key to simplifying the spacetime's CS theory in the following section.

\section{Non-orientable $\ads$'s CS/WZW correspondence} %
\label{sec:correspondence}

\subsection{CS theory of orientable $\ads$}
\label{sec:CS}

Einstein's equations can be derived from extremizing the Einstein-Hilbert action,
  \begin{equation}
    S_{\text{EH}}[g] \equiv \frac{1}{16 \pi G} \int_{\mathcal M} \text{d}^3 x \sqrt{-g} (R - 2 \Lambda) + \Gamma,
  \end{equation}
  where $\Gamma$ is a boundary term.

  This can be written in terms of its first-order (or Palatini) formulation~\cite{Misner73} by defining a vielbein in terms of the  ``square root'' of the metric, $g_{\mu \nu}(x) \equiv e^a_\mu(x) \eta_{a b} e^b_\nu(x)$, and a spin connection $\omega^a_b$ such that $T^a \equiv \text d e^a + \omega^a_b \wedge e^b$ is covariant under the isometries of the spacetime:
  \begin{equation}
    S_{\text{EH}}[e,\omega] = \frac{1}{16 \pi G} \int_{\mathcal M} \left( e^a \wedge R_a[\omega] - \frac{\Gamma}{3} \epsilon_{abc} e^a \wedge e^b \wedge e^c \right).
  \end{equation}

This was shown to be similar to the CS theory action in three dimensions~\cite{Witten88,Donnay16},
  \begin{equation}
    S_{\text{CS}}[A] = \frac{k}{4 \pi} \int_{\mathcal M} \Tr \left( A \wedge \text d A + \frac{2}{3}A \wedge A \wedge A \right),
  \end{equation}
  where, given a Poincar\'e symmetry group, $A_\mu$ is the gauge generated by its translations $P_a$ and Lorentz transformation $J_{ab}$ with weights given by the vielbein and spin connection:
  \begin{equation}
    A_\mu \equiv \frac{1}{\ell} e^a_\mu P_a + \omega^a_\mu J^a.
  \end{equation}
  
  For $\ads$, $\Lambda < 0$ and the gauge group is $\text{SO}(2,2)$, 
which relates the Einstein-Hilbert to the CS action when $\Lambda = -1/\ell^2$ and $k=\ell/(4 G)$. 

  Redefining,
  \begin{equation}
    J^\pm_a \equiv \frac{1}{2}(J_a \pm P_a),
  \end{equation}
it follows that two independent gauges can be defined,
  \begin{equation}
    A^+ \equiv (e^a/\ell + \omega^a) T_a, \quad A^- \equiv (e^a/\ell - \omega^a) T_a,
  \end{equation}
  with $T_a$ defined as generators of $\mathfrak{sl}(2, \mathbb R)$. The Chern-Simons action can be written as an antilinear combination of these two chiral CS actions:
  \begin{equation}
    S_{\text{CS}}[\Gamma] = S_{\text{CS}}[A^+] - S_{\text{CS}}[A^-].
  \end{equation}

  The linear combination of these two chiral CS actions is also valid and corresponds to the Chern-Simons gravity term from another non-degenerate invariant bilinear form~\cite{Witten88}. Classically, these two actions are indistinguishable because they produce the same Einstein equations. In general, this implies that these two actions are independent of each other after quantization as well, which leads to the problems classically interpreting tensored ground and excited states that were discussed in section~\ref{sec:intro}. 

\subsection{BF theory extension}
\label{sec:BFandCS}

CS theory is limited to gauge fields $A_\mu$ that map from orientable manifolds since it involves an integral of a $3$-form. Generalizing such an integral to handle non-orientability requires the inclusion of $3$-form densities, which obtain a minus sign under a coordinate transformation with a negative Jacobian determinant~\cite{Chen14}.

Prior work has shown that such an extension, which involves defining $3$-form densities in four-dimensional BF theory on the non-orientable manifold $\mathcal M$, is equivalent to CS theory on the orientable double cover manifold $\widetilde{\mathcal M}$~\cite{Chen14}. This equivalence takes regular forms and density forms on $\mathcal M$ to so-called even ($+$) and odd ($-$) forms on $\widetilde{\mathcal M}$, respectively, which are defined by their action under the Mobius twist (eq.~\ref{eq:Mobiustwist}) acting on the double cover, $\sigma_Z: \widetilde{\mathcal M} \rightarrow \widetilde{\mathcal M}$, 
\begin{equation}
  \Omega^p(\widetilde{\mathcal M}) = \Omega^p_+(\mathcal M) \oplus \Omega^p_-(\widetilde{\mathcal M}),
\end{equation}
with
\begin{equation}
  (\sigma_Z^* \tilde\omega_\pm)(\tilde x) = \pm\tilde \omega_\pm(\tilde x),
\end{equation}
where $\sigma^*_Z$ is the pullback of the Mobius twist. In this way, this action splits $p$-forms into even and odd forms.

Defining $\widetilde A$ to be the even form on $\widetilde{\mathcal M}$ and $\widetilde B$ the odd form, we can define $A^\pm = \widetilde A \pm \ell^{-1} \widetilde B$, it follows that the BF action $I_{\text{BF}}$ is related to the CS action $I_{\text{CS}}$ by:
\begin{equation}
  I[\widetilde A, \widetilde B; \widetilde M]_{\text{BF}} = \frac{1}{2} \left( I[A^+; \widetilde M]_{\text{CS}} - I[A^+; \widetilde M]_{\text{CS}} \right),
\end{equation}
where
\begin{equation}
  I[A^\pm; \widetilde M]_{\text{CS}} = \frac{k_{\text{CS}}}{4 \pi} \int_{\widetilde M} \Tr \left( A^\pm \wedge d A^\pm + \frac{2}{3} A^\pm \wedge A^\pm \wedge A^\pm \right),
\end{equation}
for $k_{\text{CS}} = \ell k_{\text{BF}}/2$ the coupling constant in CS theory~\cite{Chen14}.

Since $\widetilde A$ is an even one-form field and $\widetilde B$ is an odd one-form field, it follows that $\sigma^*_Z(A^+) = A^-$. It therefore follows that $I[A^+; \widetilde M]_{\text{CS}} = - I[A^-; \widetilde M]_{\text{CS}}$ and so $  I[e, \omega; M]_{\text{BF}} = \pm I[A^\pm; \widetilde M]_{\text{CS}}$~\cite{Chen14}. 
This agrees with our observation that the $\mathbb Z_2$ quotient defining non-orientable $\spacetime$ identifies the left- and right-handed $\slr$ isometry factors with each other via the mirror symmetry that identifies clockwise with anti-clockwise twisted Mobius strips. 

As a consequence, the Chern-Simons action $I[A^+; \widetilde M]_{\text{CS}} + I[A^-; \widetilde M]_{\text{CS}} = 0$ and only the standard Einstein-Hilbert action is possible from the CS theory. This is because there is only independent chiral CS action in the non-orientable case. As we shall discuss in section~\ref{sec:discussion}, this greatly simplifies the classical interpretation of allowed excited states.

The Brown-Henneaux boundary conditions~\cite{Brown86} that reveal the existence of a central extension to $\ads$'s asymptotic isometries allow for diffeomorphisms that change the orientation in the bulk $\ads$ with respect to a fixed conformal boundary and vice-versa~\cite{Loran10}. However, choosing a sign for energy and angular momentum on a boundary CFT$_2$ fixes the boundary orientation and so prevents the latter set of diffeomorphisms from consideration~\cite{Loran10}. As we have seen in Section~\ref{sec:construction}, while the diffeomorphism that defines our non-orientable $\ads$ changes the orientation of the boundary compared to the bulk double cover of $\ads$, this means that from the perspective of (a single cover) of $\ads$, it changes the orientation of the bulk with respect to a fixed conformal boundary, as desired. This is why in section~\ref{sec:constructionastwopatches} when we looked at the Poincar\'e patches of $\spacetime$ we found that the twist was located at their glued together boundaries in the bulk, while in section~\ref{sec:adsisometries} when we looked at the double cover we found the twist (fixed point) at the boundary.

Determining the central charge via the Brown-Henneaux method only requires examining the asymptotic behavior of diffeomorphisms, which means that it trivially applies to analyzing this spacetime and implies a Virasoro algebra with central charge. However, here it implies that the central charge of the left- and right-handed isometries must be the same since these are equivalent isometries and not independent of each other. This agrees with prior work that established an algebraic relationship between an orientation-free AdS$_3$ and a single central charge~\cite{Kovalsky23}.

Finally, we note that the single CS term reduces to a single chiral WZW model, i.e.~a vertex operator algebra, at its boundary. This has been shown~\cite{Coussaert95,Donnay16} by first adding a surface term $-\kappa \int_{\partial \mathcal M} \dee \tau \dee \phi \Tr ( {A^+}^2_\phi + {A^-}^2_\phi )$ to the action so that the chiral part of the action becomes,
\begin{equation}
I[A^+; \widetilde M]_{\text{CS}} = I[A^+; \widetilde M]_{\text{CS}} - \kappa \int_{\partial \mathcal M} \dee \tau \dee \phi \Tr ({A^+}^2_\phi),
\end{equation}
which can then be found to reduce to 
\begin{equation}
  I[A^+; \widetilde M]_{\text{CS}} = 2 \kappa \int_{\partial \mathcal M} \dee \tau \dee \phi \Tr (g^{-1} \partial_\phi g g^{-1} \partial_- g) + \kappa \Gamma(G),
\end{equation}
where $G$ is an $\slr$ group element and $\Gamma$ is the bulk Wess-Zumino term~\cite{Novikov82,Witten84}. This is equal to a chiral WZW action for the group element $g$.

As a result, orientable $\ads$'s CS action corresponds to an antilinear combination of a chiral and anti-chiral WZW term whereas $\spacetime$'s possesses just one independent term. This means that there is no way recover a non-chiral WZW action for $\spacetime$, though this is also true for regular $\ads$ if zero modes are included~\cite{Coussaert95,Barnich13}.

\subsection{A more exact description in CS theory}
\label{sec:moreexactCS}

As shown in section~\ref{sec:correspondence}, the two $\slr$ actions of $\spacetime$ are not independent. Defining one chiral CS action equivalently defines the other. This implies that the quantization of $\spacetime$ cannot produce a tensor factor Hilbert space corresponding to left- and right-handed state factors. Instead, a state should be fully specified by its description on only one chiral half. This has a significant implication on the excited state spectrum of this spacetime.

The quotient over the given $\mathbb Z_2$ subgroup of $\text{Spin}^+(2,2)$ does not commute with the BTZ blackhole $\mathbb Z_2$ quotient, since the latter is over both factors of $\slr$. This can be seen from the fact that for $0<r_-\le r_+$ the BTZ quotient can be expressed to be over $u \pm x \sim e^{\pm 2 \pi r_+/\ell} (u \pm x)$ and $v \pm y \sim e^{\pm 2 \pi r_+/\ell} (v \pm y)$, while for $r_-=r_+=0$ it is over $J_{23}-J_{13}$~\cite{Banados93}, neither of which can be made commuting with $u \pm x \sim x \mp u$ and $v \pm y \sim y \mp v$. It follows that these states are not valid excited states for $\spacetime$.

On the other hand, $\mathbb Z$ quotients of a single left-~or right-handed $\slr$ factor should correspond to allowed excited states. The asymptotic behavior of such $\ads$ orbifolds differs from regular $\ads$ in that their universal cover's compact direction at their boundary is lightlike, instead of timelike~\cite{Coussaert98}. While this brings into question their validity as excited states for $\ads$ without additional quantum correction~\cite{Witten07}, it makes them perfectly suitable excited states for $\spacetime$ since we showed in section~\ref{sec:constructionastwopatches} that $\spacetime$ also possesses a lightlike compact direction perpendicular to its boundary.

This significantly simplifies the excited state spectrum of $\spacetime$ to include states corresponding to orbifolds of either $\slr$ factor of regular $\ads$.
Such states in the negative sector of the BTZ's mass spectrum are easily described by CS theory~\cite{Mivskovic09}. In particular, they correspond to a coupling between $2{+}1$ gravity and external currents produces by point sources with ``negative mass'', which partially break gauge invariance based on the specific fixed point and direction they point at in the spacetime. While the field equations in defect-free $\ads$ are $F =0$, the point sources produce $F = j$ with $j \sim q \delta (T) \text d \Omega_G$. 
In other words, point source defects have an associated finite number of degrees of freedom instead of BTZ black holes' infinitely many.

We note that unitarity is still preserved in the corresponding boundary WZW model of $\spacetime$'s CS theory, despite it not possessing a decomposition into independent $\slr$ tensor factors. The boundary CFT$_2$ of $\ads$ requires the difference in central charges associated with the two factors to take certain values in order to be unitary~\cite{Witten07}. This is trivially satisfied when they two actions are equivalent (up to a minus sign), as we demonstrated in section~\ref{sec:BFandCS} for $\spacetime$. However, the central charge itself still needs to take on certain values. There are some tantalizing hints from the compact gauge case that non-orientability by itself may impose similar constraints and thereby explain such discretization~\cite{Chen14}, though we leave this for future work. 
  
\section{Discussion}
\label{sec:discussion}

As mentioned in section~\ref{sec:construction}, $\spacetime$ is not an $\ads$ orientifold like existing non-orientable $\ads$ manifolds. These prior constructions have taken $\mathbb Z_2$ quotients of $\text{O}(2,2)$ to break orientation instead of its spinorial double cover. In many respects, this means that $\spacetime$ is merely the non-orientable version of $\ads$ (much like a Mobius strip is the non-orientable version of a cylinder, and $\mathbb{RP}^n$ is of $\mathbb R^n$, etc.)~because it has a very similar isometry structure to regular $\ads$. 
Both spacetimes consist of local patches with six $\mathfrak{sl}(2,\mathbb R) \oplus \mathfrak{sl}(2,\mathbb R)$ isometries, but are just glued together differently so that $\spacetime$ is globally non-orientable while $\ads$ is orientable. As a result of the different requirements of non-orientable glueing, $\spacetime$ obtains a compact lightlike direction at its boundary while $\ads$ obtains a spacelike direction.

It can be instructive to compare the $\mathbb Z_2$ subgroup that defines the quotient of $\spacetime$ (corresponding to $u$-$x$ and $v$-$y$ exchange) with the exchange of $u$-$v$ and $x$-$y$ that defines one example of an orientifold of $\ads$~\cite{Loran10}. Both $\mathbb Z_2$ are defined in terms of transformations in $\rtt$. In the latter case, this is a symmetry of $\text{O}(2,2)$ and not of $\text{SO}(2,2)$ and produces an opposite orientation at the boundary of $\ads$ compared to its bulk. In the former case, this is instead a symmetry of $\text{Spin}^+_R(2,2)$ and not of $\text{SO}(2,2)$, and as result produces an opposite orientation in the bulk of $\ads$ compared to its boundary.

As a result of this spacetime's fixed boundary orientation, a CFT$_2$ supported on this boundary can be defined with a chosen sign for its energy and angular momentum, which permits a more natural comparison between its bulk and boundary. For instance, $\ads$ orientifolds with changed bulk orientation with respect to their boundary have been found to be associated with well-defined operators in the boundary CFT$_2$~\cite{Loran10,Loran11}. However, this ceases to be the case when the orientifold fixed point is lightlike, where they become solutions to the Einstein equations everywhere. This latter case holds here due to the lightlike boundary of the double cover which inherits a singularity after the $\mathbb Z_2$ quotient, and which becomes a bulk lightlike fixed point from the perspective of its non-orientable single cover. This agrees with what we observe too, since instead of producing a well-defined ``mass'' or central operator, we have found that our CFT$_2$ is instead more heavily restricted to have only one independent chiral factor.

Relatedly, we found that our solution $\spacetime$ has a continuous Ricci scalar at every point in the bulk and its asymptote. This must continue to hold for its orbifold excited states, since this is the case for their orientable counterparts found in~\cite{Coussaert98}. As mentioned, their $\mathbb Z$ quotients commute with our $\mathbb Z_2$ quotient so this property will be preserved when taken together. A similar observation was made of the orientifolds of these excited states when they produced lightlike fixed points~\cite{Loran11}. This is in contrast to the general case of orbifolds of $\ads$, which inherit fixed points where their Ricci scalar is divergent. When this is not the case, 
corresponding orbifolds satisfy Einstein's equations at every point of their manifold. 

This brings us to speculate on an interesting reason that CS theory might be able to exactly describe non-orientable $\ads$ gravity, at least perturbatively: it is only in this case that the ground state, corresponding to $\spacetime$, and its self-dual excited states with a lightlike compact direction, satisfy Einstein's equations \emph{at every point on their manifold} (even on their orbifold fixed point). Given that CS theory's only input are the local isometries without reference to a metric, it seems reasonable that it would only exactly describe a gravity theory that satisfies its equations everywhere (i.e.~is truly metric-agnostic).

The $\spacetime$ solution seems to be a physically realistic spacetime. In particular, its closed timelike curves are removable, and due to its orientation-breaking $\mathbb Z_2$ quotient over $\text{Spin}^+_R(2,2)$, it is able to sidestep a spinorial interpretation that can arise if the gauge group of CS theory is actually chosen to be $\text{Spin}^+(2,2) \sim \slr \times \slr$ (or its off-diagonal variants). Two-dimensional representations of $\slr$ produce two-dimensional real vector bundles that generalize spin bundles in classical geometry, which are more naturally associated with fermions instead of pure gravity~\cite{Witten07}. In this sense, the orientability-breaking quotient can thought of as the step that prevents this association.

Finally, we note that perhaps it is not surprising that a restriction to non-orientability produces a more exact correspondence with CS theory. Without BTZ black holes, the corresponding excited state spectrum is so significantly more trivial compared to orientable $\ads$, that it may perhaps not be physically interesting. On the other hand, exact examples of the holographic principle are uncommon enough to be valuable and we hope this one can still be useful in elucidating the general case. Perhaps this restricted example can be generalized to scenarios with more degrees of freedom.

As a final remark, we note that some aspects of this $\spacetime$ spacetime seem to overlap with properties that are observed in the so-called ``chiral'' $\ads$ spacetime~\cite{Li08,Strominger08}, namely the dependence of its boundary CFT$_2$ on only one chiral component. However, chiral $\ads$ spacetimes still possess their full $\slr\times \slr$ isometries classically and lose one factor's central extension upon quantization. As a result, chiral $\ads$ is generally regarded to be orientable and thus is a very different spacetime from $\spacetime$.

\section{Conclusion}
\label{sec:conc}

We derived a new non-orientable $\ads$ solution to three-dimensional gravity with negative cosmological constant. Unlike existing solutions, which are ``orientifolds'' of regular $\ads$, our solution is instead an orientifold of its double cover. We showed that it is a valid classical approximation to an $\ads$ ground state, and that it exhibits a simplified excited state spectrum lacking BTZ black holes. As a result, we argued that it is more exactly described by CS theory (more specifically, by its non-orientable generalization). We noted that, unlike in the orientable case, the classical geometries corresponding to this spectrum of states are all exact solutions to Einstein's equations at every point on their manifold.

\appendix

\section{Identifying the $\text{Spin}^+_R(2,2)$ isometry group}
\label{sec:spindoublecover}

Consider the geometry
\begin{equation}
  \text d s^2 = \text d (a^2 + b) \text d d - \text d (2 a b) \text d c = \det (\sum \frac{\text d M'}{\text d x_\mu}\text d x^\mu),
\end{equation}
where $a$, $b$, $c$, $d \in \mathbb R$ and 
\begin{equation}
M' \equiv \begin{pmatrix} a^2 - b^2& 2 a b\\ c & d \end{pmatrix}.  
\end{equation}
Similarly to the metric for $\rtt$, this is invariant under transformations $(S_L, S_R) \in (\slr \times \slr)/\mathbb Z_2$: $S_L^T M' S_R$.

However, unlike regular $\rtt$, these transformations only affect the ``external'' degrees of freedom $a^2 - b^2$, $2 a b$ and are ambiguous about the ``internal'' degrees of freedom $a$ and $b$. For instance, consider the case $S_L = I$ and $S_R = \begin{pmatrix}s_{11} & s_{12}\\ s_{21} & s_{22}\end{pmatrix}$:
\begin{align}
  (a')^2 - (b')^2 &= s_{11} (a^2 - b^2) + s_{12}(2 a b),\\
  2 a' b' &= s_{21}(a^2 - b^2) + s_{22}(2 a b),\\
  c' &= s_{11} c + s_{12} d,\\
  d' &= s_{21} c + s_{22} d,
\end{align}
where $s_{11} s_{22} - s_{12}s_{21} = 1$. There is no way to determine $(a',b')$ uniquely from the above system of equations since both $(a',b')$ and $(-a',-b')$ satisfy any solution.

We can recognize this degeneracy of solutions as an added symmetry in this spacetime: $\sigma_Z:(a,b,c,d) \rightarrow (-a,-b,c,d)$. This $\mathbb Z_2$ symmetry group can be included in a continuous set of isometries more easily understood in terms of complex coordinates: $z \equiv a + i b$ and $w = c+ id$. Consider the isometry transformation
\begin{align}
  \Omega(\omega) \circ (z,w) &\rightarrow (z e^{i \omega}, w e^{2 i \omega}),\\
  \implies (a,b) &\underset{{\Omega(\omega)}}{\rightarrow} (a \cos\omega + b \sin \omega, b \cos \omega - a \sin \omega),\\
  (c,d) &\underset{{\Omega(\omega)}}{\rightarrow} (c \cos2\omega + d \sin 2\omega, c \cos 2\omega - d \sin 2\omega).
\end{align}
$\Omega(\omega = \pi)$ corresponds to the $\sigma_Z$ transformation.

\begin{widetext}
In terms of the external coordinates
\begin{equation}
  \begin{pmatrix} a^2-b^2 & 2ab\\ c & d \end{pmatrix} \underset{{\Omega(\omega)}}{\rightarrow} \begin{pmatrix} (a^2-b^2)\cos(2\omega)+(2ab)\sin(2\omega) & (2ab)\cos(2\omega)-(a^2-b^2)\sin(2\omega)\\ c\cos(2\omega)+d\sin(2\omega) & d\cos(2\omega)-c\sin(2\omega) \end{pmatrix},
\end{equation}
which is the same as a right-handed rotation matrix with angle $2 \omega$. However, in the internal coordinates, $\omega$ and $\omega+\pi$ correspond to $(a,b)$ and $(-a,-b)$, respectively, and are distinguishable. As a result, $\Omega(\omega)$ double covers the right-handed $\text{SO}(2) \subset \slr \subset \text{Spin}^+_R(2,2)$.
\end{widetext}

Note that the left-handed $\slr$ isometry does not become double covered, as befitting $\text{Spin}^+_R(2,2)$. This is because the left-hand $\slr$  symmetry mixes $(a^2-b^2)$, $c$ with $2ab$, $d$, and so there is no corresponding continuous ``internal'' $\text{SO}(2)$ symmetry for $a$ and $b$ on the left-hand side.

Formally, we can identify $\text{Spin}^+_R(2,2)$ as the double cover we are dealing with by identifying the central elements in this spacetime and comparing them to the central elements from before. Here, they are
\begin{align}
  &Z\left(\frac{\slr \times \mplr}{\mathbb Z_2}\right) \nonumber\\
  &= \{\Omega(0), \Omega(\pi), \Omega(\pi/2), \Omega(-\pi/2)\}.
\end{align}
The first two correspond to the former $I \times I$ and $-I \times -I$ elements of $\text{SO}^+(2,2) \sim (\slr \times \slr)/\mathbb Z_2$. 

\section{Computing tangent and cotangent basis transformations}
\label{sec:tangent-transformations}
To compute the transformations of the tangent and co-tangent bases, we need $a, b$ in terms of $A, B$ or equivalently the ``realification'' of $z^{1/2}$:
\begin{equation}
a+ib = \sqrt{\frac{|Z|+A}{2}} + i\frac{B}{|B|}\sqrt{\frac{|Z|-A}{2}}.
\end{equation}

Taking the appropriate partial derivatives of the above expression, the tangent vectors transform as:
\begin{equation}
\begin{pmatrix} \partial_A \\ \partial_B \end{pmatrix} = \frac{1}{2|Z|} \begin{pmatrix} \sqrt{|Z|+A} & \frac{-B}{|B|}\sqrt{|Z| - A} \\ \frac{B}{\sqrt{(|Z|+A)}} & \frac{B^4}{|B|^3} \frac{1}{\sqrt{(|Z|-A)}} \end{pmatrix} \cdot \begin{pmatrix} \partial_a \\ \partial_b \end{pmatrix}.
\end{equation}
The cotangent vectors transform as:
\begin{align}
  \begin{pmatrix} \dee A \\ \dee B \end{pmatrix} &= \begin{pmatrix} 2a & -2b \\ 2b & 2a \end{pmatrix} \cdot \begin{pmatrix} \dee a \\ \dee b \end{pmatrix}\\
  &= \begin{pmatrix} \sqrt{|Z|+A} & -\frac{B}{|B|}\sqrt{|Z|-A} \\ \frac{B}{|B|}\sqrt{|Z|-A} & \sqrt{|Z|+A} \end{pmatrix} \cdot \begin{pmatrix} \dee a \\ \dee b \end{pmatrix}.
\end{align}
We can also check that
\begin{align}
& \dee A \circ \partial_A = 1, \ \dee A \circ \partial_B = 0, \\
& \dee B \circ \partial_A = 0, \ \dee B \circ \partial_B = 1.
\end{align}

To understand these results a little more carefully, we can use polar coordinates to parameterize the coordinate $Z = A+iB = R(\cos\Theta + i \sin\Theta)$ where $\Theta \in (-\pi, \pi)$.
The result is:
\begin{equation}
\begin{pmatrix} \partial_A \\ \partial_B \end{pmatrix} = \frac{1}{2\sqrt{R}} \begin{pmatrix} \cos\frac{\Theta}{2} & -\sin\frac{\Theta}{2} \\ \sin\frac{\Theta}{2} & \cos\frac{\Theta}{2} \end{pmatrix} \cdot \begin{pmatrix} \partial_a \\ \partial_b \end{pmatrix}.
\end{equation}
For the cotangent vectors:
\begin{equation}
\begin{pmatrix} \dee A \\ \dee B \end{pmatrix} = \sqrt{R} \begin{pmatrix} \cos\frac{\Theta}{2} & -\sin\frac{\Theta}{2} \\ \sin\frac{\Theta}{2} & \cos\frac{\Theta}{2} \end{pmatrix} \cdot \begin{pmatrix} \dee a \\ \dee b \end{pmatrix}.
\end{equation}

We can also compute the Lie bracket between $\partial_A, \partial_B$ using the transformations above.
To do so, we start by writing everything in terms of $a, b$ coordinates:
\begin{align}
\begin{pmatrix} \partial_A \\ \partial_B \end{pmatrix} = \frac{1}{2(a^2+b^2)} \begin{pmatrix} a & -b \\ b & a\end{pmatrix} \begin{pmatrix} \partial_a \\ \partial_b \end{pmatrix}.
\end{align}
The commutator is then:
\begin{align}
[\partial_A, \partial_B] = \left[\frac{1}{2(a^2+b^2)}(a\partial_a - b\partial_b), \frac{1}{2(a^2+b^2)}(b \partial_a + a \partial_b) \right]. 
\end{align}
Since $a, b$ are taken as (holonomic) coordinates, we know that $[\partial_a, \partial_b] = 0$.
\begin{widetext}
The four individual terms are:
\begin{align}
& \left[\frac{1}{2(a^2+b^2)}a\partial_a, \frac{1}{2(a^2+b^2)}b\partial_a\right] = \frac{-b}{4(a^2+b^2)^2} \partial_a \\
& \left[\frac{-1}{2(a^2+b^2)}b\partial_b, \frac{1}{2(a^2+b^2)}a\partial_b\right] = \frac{a}{4(a^2+b^2)^2}\partial_b \\
& \left[\frac{1}{2(a^2+b^2)}a\partial_a, \frac{1}{2(a^2+b^2)}a\partial_b\right] = \frac{a(b^2-a^2)}{4(a^2+b^2)^3}\partial_b + \frac{2a^2b}{4(a^2+b^2)^3} \partial_a \\
& \left[\frac{-1}{2(a^2+b^2)}b\partial_b, \frac{1}{2(a^2+b^2)}b\partial_a\right] = \frac{b(b^2-a^2)}{4(a^2+b^2)^3}\partial_a - \frac{2b^2a}{4(a^2+b^2)^3} \partial_a
\end{align}
As such, we find that $[\partial_A, \partial_B] = 0$.
\end{widetext}

\noindent ---
L.~K.~Kovalsky~thanks A.~Dhumuntarao and Y.~W.~Fan for their helpful discussions during this study.

This material is based upon work supported by the U.S. Department of Energy, Office of Science, Office of Advanced Scientific Computing Research, under the Quantum Testbed Pathfinder program.

Sandia National Laboratories is a multimission laboratory managed and operated by National Technology \& Engineering Solutions of Sandia, LLC, a wholly owned subsidiary of Honeywell International Inc., for the U.S. Department of Energy’s National Nuclear Security Administration under contract DE-NA0003525. 
This paper describes objective technical results and analysis. 
Any subjective views or opinions that might be expressed in the paper do not necessarily represent the views of the U.S. Department of Energy or the United States Government.

\bibliography{biblio}
\bibliographystyle{unsrt}

\end{document}